\documentclass[aps,preprint,sort&compress,square,floatfix]{revtex4}
\usepackage[dvips]{graphicx}

\begin{document}

\title{Distorted Waves with Exact Non-Local Exchange: a Canonical Function Approach}

\author{K. Fakhreddine \dag, R. J. Tweed \ddag, G. Nguyen Vien
\ddag, C. Tannous \S, J. Langlois \ddag, and O. Robaux \ddag}

\affiliation{\dag Faculty of Science, Lebanese University and C.N.R.S. of Lebanon, 
P.O. Box: 113-6546, Beirut, Lebanon \\
\ddag Laboratoire de Collisions Electroniques et Atomiques, U.F.R. Sciences et 
Techniques, 6 avenue Le Gorgeu BP: 809, 29285 Brest Cedex, France \\
\S  Laboratoire de Magn\'etisme de Bretagne, U.M.R. 6135 du
C.N.R.S., U.F.R. Sciences et Techniques, 6 avenue Le Gorgeu BP: 809, 29285 Brest
Cedex, France}

\begin{abstract} 
It is shown how the Canonical Function approach can
be used to obtain accurate solutions for the distorted wave problem taking
account of direct static and polarisation potentials and exact non-local
exchange. Calculations are made for electrons in the field of atomic hydrogen
and the phaseshifts are compared with those obtained using a modified form of
the DWPO code of McDowell and collaborators: for small wavenumbers our 
approach avoids numerical instabilities otherwise present. 
Comparison is also made with
phaseshifts calculated using local equivalent-exchange potentials and it is
found that these are inaccurate at small wavenumbers. Extension of our method
to the case of atoms having other than s-type outer shells is dicussed.
\end{abstract}

\maketitle

\section{Introduction}

Distorted Wave Born calculations of Triple Differential Cross Sections (TDCS)
for electron impact ionization of an atom or ion require the resolution of
the integro-differential equations satisfied by the radial parts of the free
electron wavefunctions. Very often a simplifying approximation is made
through replacement of the exchange operator by a central equivalent exchange
potential (Furness and McCarthy \cite{Furness}, Bransden and Noble \cite{Bransden}). But while
this is probably satisfactory for electrons with energies of a few eV and
more it may cause problems near to threshold. Furthermore, to interpret the
new generation of experiments using polarized electrons the proper inclusion
of exchange may well be important. There exists a means of exact solution of
the equation for the radial wavefunction including static and static-exchange
terms (the diagonal parts of the direct potential and the exchange operator)
and a polarisation potential. This is the Distorted Wave Polarised Orbital
(DWPO) method of McDowell \textit{et al} \cite{McDowella} which was developed for Distorted
Wave Born calculations of excitation of Hydrogen (McDowell \textit{et al} \cite{McDowellb}) and
of Helium \cite{Scott} and replaces the integro-differential
equation by coupled differential equations. In both cases, only an s-state of
the atom is considered and the central potentials are expressed in analytical
form.

The phase shift and the wavefunction in the asymptotic region were determined
by use of the analytic second order JWKB solution (Burgess \cite{Burgess}). If the
energy of the free electron is $k^2$ Ry the integro-differential
equation can be rewritten in terms of the variable $k r$ but the
value of the radius out to which exchange and the short-range part of the 
static potential remain significant is determined by the extent of the electron 
cloud of the atom. For small $k$ the second order JWKB solution is valid only 
at radii very much larger than 
that of the onset of the asymptotic region where exchange and short-range 
potentials are negligible. So we have modified the published McDowell 
\textit{et al} code by using the iterative numerical JWKB code developped 
by Klapisch, Robaux and collaborators (see \cite{bar}), which is valid 
throughout the asymptotic region. The solutions are
started by means of series expansions at the origin and continued by Numerov
integration: the form of series corresponding to the regular solution is
imposed by taking a power $r^{l' + 1}$ at the origin in the $l'$
partial wave. For small $k$ the choice of integration step and
changeover point is delicate, since convergence of the series
requires small r whereas the Numerov integration becomes unstable (picking up
some of the irregular solution) if it is started at too small a radius. In
spite of modifications to the original code we have found it hard to obtain a
solution which is stable in the sense of the phaseshift being independent of
a $20 \%$ change in steplength to better than one significant figure at $k=
0.01$ and two significant figures at $k= 0.1$. But a similar but
simpler code using equivalent-exchange potentials in a single differential
equation is numerically stable.

The aim of the present work is to develop an alternative numerical method,
based on the Canonical Function approach (Kobeissi \textit{et al} \cite{Kobeissi88,Kobeissi91,Kobeissi91a}, which overcomes the
problems of numerical stability. In this approach two independent sets of
solutions are generated starting at some central point and integrating
inwards to the origin and outwards to the asymptotic region. Both contain
linear combinations of the regular and the irregular solutions, but by
suitably combining the two the irregular solution is eliminated. Integration
can be made out to a very large radius, allowing the phase-shift to be
determined by matching to plane or Coulomb wave solutions. It is not
necessary to obtain a series expansion of the solution to start the
integration, which makes it unnecessary to have a series expansion of the
potential. Our method is therefore convenient when potentials are generated
numerically rather than analytically, although here we have worked with the
analytical potentials of McDowell and collaborators for the sake of
comparison. Even in the case of small atoms, this is an advantage if
numerical polarization-correlation potentials obtained from the density
functional theory are used, although the series expansion of these near to
the origin can be generated by a polynomial fit. Furthermore we can very
easily extend our work to the case of p-state and d-state atoms whereas
defining series starting solutions in these cases is difficult to do even on
a case-by-case basis. The added complexity, with a much larger number of
coupled equations to solve, would in any case probably cause even worse
numerical instability at low $k$.

It is worthwhile developing a version of the DWBA approximation including
non-local exchange and applicable to heavy atom targets. This is because the
very successful Converged Close Coupling (CCC) approach of Bray \cite{Bray}
becomes increasingly difficult to apply as target complexity increases. In
section 2 we discuss our present implementation of the DWPO method for the
non-local exchange problem, which we propose to use to generate the necessary
wavefunctions. We adopt essentially the notation of Rouet \textit{et al} \cite{Rouet}: a
full description of the DWPO method is given in the appendix ot the latter paper.
In section 3 we describe the use of the the Canonical Fuction approach to
solve the resulting coupled differential equations. It is interesting to note
that the present treatment of exchange can ultimately be combined with the
Canonical Function approach to the solution of coupled equations without
exchange \cite{Fakhreddine94,Fakhreddine99,Kobeissi91a} so as to get a full solution of the Close Coupling
equations including all potentials and non-local exchange terms (diagonal and
off-diagonal). In section 4 we compare the phase shifts obtained by our new
code with those given by our modified form of the McDowell code and by
solution of single differential equations with local equivalent-exchange
potentials. We highlight both the inaccuracies of the McDowell code, even in
its modified form, and the indequacies of local exchange approximations.

\section{The integro-differential equation}

We consider the impact of a free electron of energy $k^2$ Ry on
a one-electron atomic system of nuclear charge Z in a 1s atomic state of
energy $E_{10}$ Ry with a radial wavefunction $R_{10} ( r ) = 2 Z^
{3 / 2} r \exp ( - Z r)$. For a free electron with angular
momentum quantum numbers ${l' , m'}$, if we include only on-diagonal
potentials and exchange operators and replace the neglected off-diagonal
coupling potentials by a polarisation potential $V_{pol} ( r )$,
its radial wavefunction $F_{l'} ( k , r )$ satisfies the
intego-differential equation:
\begin{equation}
\left[ {\frac{\partial^2}{\partial r^2}} -{\frac{l' ( l' + 1
)}{r^2}} + k^2 + V_{1 s} ( r ) + V_{\mathrm{pol}\;} ( r )
+ W_{l'} ( r ) \right] F_{l'} ( k , r ) = 0
\end{equation}
where

\begin{displaymath}
V_{1 s} ( r ) ={\frac{2 Z}{r}} - {\frac{2}{r}}{\int^r_0} \left| R_{10} ( r'
) \right|^{2} d r' - {\int^{\infty}_r}{\frac{2}{r'}} \left| R_{10} ( r' ) \right|^{2} d r' 
\end{displaymath}
\begin{eqnarray*}
W_{l'} ( r ) F_{l'} ( k, r ) = ( - 1 )^{S + 1} R_{10} ( r )
\left\{ \left[ E_{10} - k^2 \right] \delta_{l' , 0} {\int^{\infty}_0} R_{10}
( r' ) F_{l'} ( k , r' ) d r'\right. \\
\left. - {\frac{2}{r}}{\int^r_0} R_{10} ( r' ) F_{l'} ( k , r' ) d r' -
{\int^{\infty}_r} {\frac{2}{r'}} R_{10} ( r' ) F_{l'} ( k , r' ) d r' \right\} 
\end{eqnarray*}

This integro-differential equation can be transformed to the following system
of coupled differential equations, which is the starting point for McDowell
\textit{et al} \cite{McDowella, McDowellb}:
\begin{eqnarray*}
{\frac{\partial^2}{\partial r^2}} F_{l'} ( k , r ) - \left[ {\frac{l'(l'
+ 1)} {r^2}} -{ \cal V}_{1 s} ( r ) \right] F_{l'} ( k ,
r ) \\ + ( - 1 )^S R_{10} ( r ) \left[ {\frac{2}{r}} {\frac{1}{2 l' + 1}} \right]
G_{l'} ( k , r )  = ( - 1 )^{S + 1} R_{10} ( r ) \delta_{l' , 0} A ( k)
\end{eqnarray*}
\begin{equation}
{\frac{\partial^2}{\partial r^2}} G_{l'} ( k, r ) -{\frac{l'(
l' + 1 )} {r^2}} G_{l'} ( k , r ) + {\frac{2 l' + 1}{r}}
R_{10} ( r ) F_{l'} ( k , r ) = 0
\end{equation}
where
\begin{equation}
A ( k ) = \left[ k^2 - E_{10} \right] {\int^{\infty}_0} R_{10} ( r' )
F_{l'} ( k , r' ) d r' 
\end{equation}
$\delta_{l' , 0}$ is the Kronecker delta and ${\cal V}_{1 s} ( r )
= k^2 + V_{1 s} ( r ) + V_{pol} ( r )$.
Our aim is to solve this system subject to the boundary conditions
\begin{equation} \begin{array}{l c r}
F_{l'} ( k , r ) \mathop{\longrightarrow}\limits_{r \rightarrow 0} 0 
& \mbox{\hspace{5.7cm}} &
G_{l'} ( k , r ) \mathop{\longrightarrow}\limits_{r \rightarrow 0} 0
\end{array} \end{equation}
\begin{equation} \begin{array}{l r}
F_{l'} ( k , r ) \mathop{\longrightarrow}\limits_{r \rightarrow \infty} a_{l'} ( k )
\left\{ s_{l'} ( k r ) - \tan \left[ \delta_{l'} ( k ) \right] c_{l'} ( k r) \right\}
&
G_{l'} ( k , r ) \mathop{\longrightarrow}\limits_{r \rightarrow \infty} 0
\end{array} \end{equation}
where $a_{l'} ( k )$ is a normalisation factor, $\delta_{l'} ( k)$ 
is the phase shift for specific $\{ k , l' \}$ and
$ s_{l'} ( \rho )$ and $c_{l'} ( \rho )$ are
respectively:
\begin{itemize}
\item
$\rho$ multiplied spherical Bessel and Neumann functions when $Z =
1$ (so that ${\cal V}_{1 s} ( r )$ is a short range potential
falling off faster than $r^{- 1}$ as $r$ tends to infinity);
\item
regular and irregular Coulomb wavefunctions when $Z > 1$
(so that ${\cal V}_{1 s} ( r )$ is a long range potential behaving
like $( Z - 1 ) r^{- 1}$ when $r$ tends to infinity).
\end{itemize}

In the more general case of an ion with a frozen core and an outer shell of
electrons in $\{n , l \}$ states of radial wavefunction
$R_{nl} ( r )$, and a free electron in the state
$\{ k , l' \}$ we have a larger set of coupled equations: 
\begin{eqnarray*}
{\frac{\partial^2}{\partial r^2}} F_{l'} ( k , r ) - \left[{\frac{l' (
l' + 1 )}{r^2}} -{\cal V}_{n l} ( r ) \right] F_{l'} (
k , r ) \\ - ( - 1 )^S R_{nl} ( r ){\frac{2}{r} \sum_{\lambda}
J_{l , l' , \lambda}} G_{l'}^{\lambda} ( k , r )
= ( - 1 )^{S + 1} R_{n l} ( r ) \delta_{l , l'} A_{n l , l'} ( k ) 
\end{eqnarray*}
\begin{displaymath}
{\frac{\partial^2}{\partial r^2}} G_{l'}^{\lambda} ( k , r )
-{\frac{\lambda ( \lambda + 1 )}{r^2}} G_{l'}^{\lambda} ( k
, r ) + {\frac{2 \lambda + 1}{r}} R_{n l} ( r )  F_{l'} ( k , r )
= 0
\end{displaymath}
subject to the boundary conditions, for all possible $\lambda$ values
\begin{displaymath} \begin{array}{lr}
F_{l'} ( k , r ) \mathop{\longrightarrow}\limits_{r \rightarrow 0} 0
&
G_{l'}^{\lambda} ( k , r ) \mathop{\longrightarrow}\limits_{r \rightarrow 0} 0
\\
F_{l'} ( k , r )  \mathop{\longrightarrow}\limits_{r \rightarrow \infty}
a_{l'} ( k ) \left\{ s_{l'} ( k r ) - \tan \left[ \delta_{l'} ( k ) \right] c_{l'}
( k r ) \right\}  
&
G_{l'}^{\lambda} (k , r ) \mathop{\longrightarrow}\limits_{r \rightarrow \infty} 0
\end{array} \end{displaymath}
Here
\begin{eqnarray*}
{\cal V}_{n l} ( r ) = k^2 + V_{nl} ( r ) \\ - 2 \sum_{\lambda} I_{l , l' ,
\lambda} \left\{ {\int^r_0}{\frac{r'^{\lambda}}{r^{\lambda + 1}}} \left| R_{n
 l} ( r' ) \right|^{2} d r'
+{\int^{\infty}_r}{\frac{r^{\lambda}}{r'^{\lambda + 1}}} \left| R_{n l} ( r'
) \right|^{2} d r' \right\} + V_{pol} ( r ) 
\end{eqnarray*}
\begin{displaymath}
A_{n l , l'} ( k ) = \left[k^2 - E_{n l} \right] {\int^{\infty}_0}
R_{n l} ( r' ) F_{l'} ( k , r' ) d r'
\end{displaymath}
$I_{l , l' , \lambda}$ and $ J_{l , l' , \lambda}$ are angular integrals which 
depend on the number of electrons in the ion outer shell and the angular 
momentum coupling scheme. $V_{nl} ( r )$ is a central potential for attraction of an
electron by the core and $E_{n l}$ is the total energy of the outer
shell electrons. This is not applicable to hydrogenic ions as the degeneracy of the
energy in $l$ makes it essential to include channel coupling potentials.
So we will not consider it further here except to note that $ J_{l , l' ,
\lambda}$ imposes the triangular rule $\left| l - l' \right| \leq
\lambda \leq l + l'$ and $l + l' + \lambda$ even. This gives an idea of 
the number of different $G_{l'}^{\lambda}$ present and so of the  extent of the 
problem we ultimately wish to solve, in the case of more complex atoms.

\section{\bf The canonical function technique for solving the DWPO equations: }

In order to facilitate the presentation it is convenient to use $f_1$ in place of 
$F_{l'}$ and $f_2$ in place of $G_{l'}$ so as to rewrite the coupled equation system (2) as a
special case of the more general system:
\begin{displaymath}
f_1 '' ( r ) + V_{11} ( r ) f_1 ( r ) + V_{12} ( r ) f_2 ( r ) = \delta_{l'
, 0} A ( k ) W_1 ( r )
\end{displaymath}
\begin{equation}
f_2 '' ( r ) + V_{22} ( r ) f_2 ( r ) + V_{21} ( r ) f_1 ( r ) = \delta_{l'
, 0} A ( k ) W_2 ( r )
\end{equation}
with
\begin{displaymath} \begin{array}{l c l}
V_{11} ( r ) = {\cal V}_{1 s} ( r ) - \displaystyle{\frac{l' ( l' + 1)}{r^2}}
& & 
V_{12} ( r ) = ( - 1 )^S R_{10} ( r ) \left[ \displaystyle{\frac{2}{r}} {\frac{1}
{2 l' + 1}} \right]
\\
V_{21} ( r ) = \displaystyle{\frac{2 l' + 1}{r}} R_{10} ( r ) 
& &  
V_{22} ( r ) = - \displaystyle{\frac{l' ( l' + 1 )}{r^2}} 
\\
W_1 ( r ) = ( - 1 )^{S + 1} R_{10} ( r ) 
& &  
W_2 ( r ) = 0
\end{array} \end{displaymath}
We will solve this system by the canonical functions method of Kobeissi and
Fakhreddine \cite{Kobeissi91a}. We can construct the general solution of equations (6)
as
\begin{eqnarray*} 
f_1 ( r ) = f_1 ( r_0 ) \alpha_{11} ( r ) + f_1 ' ( r_0 ) \beta_{11} ( r )
\\ + f_2 ( r_0 ) \alpha_{12} ( r ) + f_2 ' ( r_0 ) \beta_{12} ( r ) + \delta
_{l' , 0} A ( k ) \sigma_1 ( r )
\end{eqnarray*}
\begin{eqnarray*}
f_2 ( r ) = f_1 ( r_0 ) \alpha_{21} ( r ) + f_1 ' ( r_0 ) \beta_{21} ( r )
\\ + f_2 ( r_0 ) \alpha_{22} ( r ) + f_2 ' ( r_0 ) \beta_{22} ( r ) + \delta
_{l' , 0} A ( k ) \sigma_2 ( r )
\end{eqnarray*}
where $\left\{ \alpha_{1 j} ( r ) , \alpha_{2 j} ( r ) \right\}$ and
$\left\{ \beta_{1 j} ( r ) , \beta_{2 j} ( r ) \right\}$ are two
different pairs of independent solutions of the homogeneous system 
\begin{displaymath}
g_1 '' r ) + V_{11} ( r ) g_1 ( r ) + V_{12} ( r ) g_2 ( r ) = 0 
\end{displaymath}
\begin{equation}
g_2 '' ( r ) + V_{22} ( r ) g_2 ( r ) + V_{21} ( r ) g_1 ( r ) = 0
\end{equation}
satisfying the initial conditions at an arbitrary point $r = r_0$
\begin{equation} \begin{array}{l c r}
\alpha_{i j} ( r_0 ) = \beta_{i j} ' ( r_0 ) = \delta_{i , j}
& &
\alpha_{i j} ' ( r_0 ) = \beta_{i j} ( r_0 ) = 0
\end{array} \end{equation}
and $\{ \sigma_1 ( r ) , \sigma_2 ( r ) \}$ is a particular solution
of the inhomogeneous system
\begin{displaymath}
h_1 '' ( r ) + V_{11} ( r ) h_1 ( r ) + V_{12} ( r ) h_2 ( r ) = W_1 ( r)
\end{displaymath}
\begin{equation}
h_2 '' ( r ) + V_{22} ( r ) h_2 ( r ) + V_{21} ( r ) h_1 ( r ) = W_2 ( r )
\end{equation}
satisfying the initial conditions at $r = r_0$ 
\begin{equation}
\sigma_i ( r_0 ) = \sigma_i' ( r_0 ) = 0
\end{equation}
In matrix form: 
\begin{equation}
Y ( r ) = \alpha ( r ) Y ( r_0 ) + \beta ( r ) Y ' ( r_0 ) + \delta_{l' , 0}
A ( k ) \sigma ( r )
\end{equation}
with
\begin{displaymath} \begin{array} {l c r}
Y ( r ) = \left( \begin{array}{c} f_1 ( r )\\ f_2 ( r )\\ \end{array} \right)
& & 
\alpha ( r ) = \left\{ \begin{array}{cc} \alpha_{11} ( r ) & \alpha_{12} 
( r )\\ \alpha_{21} ( r ) & \alpha_{22} ( r )\\ \end{array} \right\}
\\ & & \\
\sigma ( r ) = \left( \begin{array}{c} \sigma_1 ( r )\\ \sigma_2 (
r )\\ \end{array} \right)
& &
\beta ( r ) = \left\{ \begin{array}{cc} \beta_{11} ( r ) & \beta_{12}
( r )\\ \beta_{21} ( r ) & \beta_{22} ( r )\\ \end{array} \right\}
\end{array} \end{displaymath}
Using the boundary conditions (4) imposes $\alpha ( 0 ) Y ( r_0 ) +
\beta ( 0 ) Y ' ( r_0 ) + \delta_{l' , 0} A ( k) \sigma ( 0 ) = 0$,
which leads to $\beta^{- 1} ( 0 ) \alpha ( 0 ) Y ( r_0 ) + Y ' ( r_0
) + \delta_{l , 0} A ( k ) \beta^{- 1} ( 0 ) \sigma ( 0 ) = 0 $ where
$\beta^{- 1} ( r )$ is the inverse of the matrix $\beta ( r )$.
Thus the constant matrices $Y ' ( r_0 ) $ and $Y ( r_0 )$ are related by:
\begin{displaymath} \begin{array}{l c r}
Y ' ( r_0 ) = Y ( r_0 ) \Lambda + \delta_{l' , 0} A ( k) \lambda
& & \\ & & \\
\Lambda = - \beta^{- 1} ( 0 ) \alpha ( 0 )
& &
\lambda = - \beta^{-1} ( 0 ) \sigma ( 0 )
\end{array} \end{displaymath}
Substituting back into (11) we then get
\begin{displaymath} \begin{array}{l c r}
Y ( r ) = \varphi ( r ) Y ( r_0 ) + \delta_{l' , 0} A ( k ) \gamma ( r)
& & \\ & & \\
\varphi ( r ) = \alpha ( r ) + \beta ( r ) \Lambda
& &
\gamma ( r ) =\beta ( r ) \lambda + \sigma ( r )
\end{array} \end{displaymath}
The solution constructed from $\varphi ( r )$ and $\gamma ( r )$ is a particular 
solution of the coupled equations (6) for which the functions 
$\left\{ f_1 (r ) , f_2 ( r ) \right\}$ are regular at the origin. It corresponds 
to initial values (where $I$ is the unit matrix) at the arbitrarily chosen 
starting point $r =r_0$ :
\begin{displaymath} \begin{array} {c c c c c c c}
\varphi ( r_0 ) = I & &
\varphi ' ( r_0 ) = \Lambda & &
\gamma ( r_0 )= 0 & &
\gamma ' ( r_0 ) = \lambda  
\end{array} \end{displaymath}

It remains to determine $A ( k )$ Substituting equations (6), for
the case $l' = 0$, into expression (3) we get
$A (k ) = \left[ k^2 - E_{10} \right] \left[ f_1 ( r_0 ) I_1 + f_2 ( r_0 )
I_2 + A ( k ) J \right]$, where
\begin{displaymath} \begin{array}{l c r}
I_1 = \displaystyle \int^{\infty}_0 R_{10} ( r ) \varphi_{11} ( r ) d r & &
I_2 = \displaystyle \int^{\infty}_0 R_{10} ( r ) \varphi_{12} ( r ) d r \\ & &
\\ J = \displaystyle \int^{\infty}_0 R_{10} ( r ) \gamma_1 ( r ) d r & &
\end{array} \end{displaymath}
This leads to
\begin{displaymath}
A ( k ) = \left[ k^2 - E_{10} \right] {\frac{ f_1 ( r_{0 )} I_1 +
f_2 ( r_0 ) I_2 }{ 1 - \left[ k^2 - E_{10} \right] J }} 
\end{displaymath}
or, in a simpler form,
\begin{equation} \begin{array}{l c r}
A ( k ) = A_1 f_1 ( r_0 ) + A_2 f_2 ( r_0 ) & & \\ & & \\
A_1=\displaystyle{\frac{\left[ k^2 - E_{10} \right] I_1}{ 1 - \left[ k^2 - E_{10}
\right] J  }} & &
A_2 =\displaystyle{\frac{\left[ k^2 - E_{10} \right]
I_2}{ 1 - \left[ k^2 - E_{10} \right] J }}
\end{array} \end{equation}
From the second of the boundry conditions (5) we have
\begin{equation}
f_1 ( r_0 ) \varphi_{21} ( r ) + f_2 ( r_0 ) \varphi_{22} ( r ) + A ( k
) \gamma_2 ( r )  \mathop{\longrightarrow} \limits_{r \rightarrow \infty} 0
\end{equation}
and comparing equations (12) and (13) we get
\begin{equation} \begin{array}{l c r}
\displaystyle{\frac{f_2 ( r_0 )}{f_1 ( r_0 )}} = D_{\infty} = \mathop{lim}
\limits_{r \rightarrow \infty} D ( r )
& &
D ( r ) = - \displaystyle{\frac{\varphi_{21} ( r ) + a_1 \gamma_2 ( r )}
{\varphi_{22} ( r ) + a_2 \gamma_2 ( r )}}
\end{array} \end{equation}
which implies, considering $f_1 ( r_0 )$ as arbitrary,  
\begin{displaymath}
f_1 ( r ) = f_1 ( r_0 )
\left\{ \left[ \varphi_{11} ( r ) + a_1 \gamma_1 ( r ) \right] + D_{\infty}
\left[ \varphi_{12} ( r ) + a_2 \gamma_1 ( r ) \right] \right\}
\end{displaymath}
From the first of boundary conditions (5) we can then determine the phaseshift
$\delta_l$ by
\begin{equation} \begin{array}{l c r}
\tan \delta_{l'} =  \mathop{lim} \limits_{r \rightarrow \infty} Q ( r ) & &
Q ( r ) = - \displaystyle{\frac{f_1 ' ( r ) s_{l'} (k r ) - f_1 ( r ) k s_{l'} ' ( k r )}
{f_1 ' ( r ) c_{l'} ( k r ) - f_1 ( r ) k c_{l'} ' ( k r )}}
\end{array} \end{equation}
We follow Kobeissi \textit{et al} \cite{Kobeissi91} in using the recursion relations for
$s_{l'} ( \rho )$ and $c_{l'} ( \rho )$ :
\begin{list}{} 
\item
$Z=1$ : 
$\displaystyle Q ( r ) = \frac{\left[ f_1 ' ( r ) -{\frac{( l' + 1 )}{r}} f_1 ( r )
\right] s_{l'} ( k r ) + k f_1 ( r ) s_{l' + 1} ( k r )}{\left[ f_1 '
( r ) -{\frac{( l' + 1 )}{r}} f_1 ( r ) \right] c_{l'} ( k r ) + k f_1
( r ) c_{l' + 1} ( k r )} $
\item
$Z>1$ : 
$\displaystyle Q ( r ) =  \frac{\left[ f_1 ' ( r ) + \left\{ {\frac{( Z - 1 )}{k ( l' +
1 )}} -{\frac{( l' + 1 )}{r}} \right\} f_1 ( r ) \right] s_{l'} ( k r ) +
\sqrt{k^2 +{\frac{( Z - 1 )^2}{( l + 1 )^2}}} f_1 ( r ) s_{l' + 1} ( k
r )}{\left[ f_1 ' ( r ) + \left\{ {\frac{( Z - 1 )}{k ( l' + 1 )}} -{\frac{(
l' + 1 )}{r}} \right\}  f_1 ( r ) \right] c_{l'} ( k r ) + \sqrt{k^2
+{\frac{( Z - 1 )^2}{( l + 1 )^2}}} f_1 ( r ) c_{l' + 1} ( k r )} $
\end{list}
The function $Q ( r )$ can be defined for any radius r. We calculate it
at large $r$ values and examine its behaviour. When it tends to a
constant limit we consider that the asymptotic region has been reached and
the phaseshift is determined to within a multiple of $2 \pi $. We
must also check that $D ( r )$ of equation (14) tends to a constant
limit when $r$ becomes large. This should happen for radii larger than
the effective extent of the atomic charge cloud, at which exchange effects
become negligible. The value of $F ( r_0 )$ is finally chosen to get the
proper normalisation of the continuum function; a $\delta$ function in
momentum requires $a_{l'} ( k ) = \sqrt{2 / \pi}$ in condition (5).

To summarise, solution of the coupled equation problem (6), without
specification of the boundary conditions, is reduced to the determination of
a set of functions having well determined initial values at some arbitray
radius $r_0$: the canonical functions given by the matrices
$\alpha ( r )$, $\beta ( r )$ and $\sigma ( r )$. We then take
linear combinations of these chosen to get other canonical functions $\varphi
 ( r )$ and $\gamma ( r )$ which satisfy the boundary condition (4)
at the origin. Finally, the asymptotic boundary conditions (5) enable us to
determine the appropriate linear combination of $\varphi ( r )$ and
$\gamma ( r )$ and to obtain the phase-shift and the wavefunction. 
Essentially, we need only to develop a single algorithm for the
solution of a system of two coupled equations of the form (7) or (9) and then
apply it to the different cases represented by the intial conditions (8) or
(10).

The enormous advantage of the present Canonical Functions approach is that the
integration of equations (7) and (9) can be started at any desired radius
$r_0$, in particular at a point far from the origin. Almost all
previous methods require a starting solution in the region near to the
origin, where numerical integration of the coupled equations cannot be
directly initiated because of the singular behaviour of the potentials which
contain terms proportional to $r^{- 1}$ and $r^{- 4}$ and the
angular momentum terms which are proportional to $r^{- 2}$. It is
possible to follow McDowell \textit{et al} \cite{McDowella,McDowellb} in obtaining a series
expansion of the regular solution at the origin, which requires the
potentials to be expressable in an analytical form and their series
expansions about the origin to be known. The solutions are continued by
numerical integration of the coupled equations using one or another of the
well-known integrators, such as the method of Numerov (1933). But for small
$k$ the outwards Numerov integration of the regular solution can
get out of control if it picks up even a very small fraction of the irregular
solution, because of ill-conditioning due to round-off errors. An alternative
is to modify the potential by introducing a hard core: ${\cal V}_{n l} (
r )$ is set artificially to infinity for $r < r_s$, where
$r_s$ is the starting point of the integration and retains its
original form for $r > r_s$. As mentioned by Bayliss \cite{Bayliss} \textit{et al} (1982), this
method gives results sensitive to the point in the classically forbidden
region at which the integration is started. If the starting point is too
small some solutions become unstable; if it is too large for the initial
conditions employed the solutions are quite simply inaccurate. In the
Canonical Functions approach, the solutions $\alpha ( r )$, $\beta ( r
)$ and $\sigma ( r )$ initially generated are in fact linear
combinations of the regular and irregular solutions of the coupled equations
and by taking the linear combinations $\varphi ( r )$ and $\gamma ( r)$ 
we eliminate the irregular solution between them.

\section{Results and Discussion}

We apply the present method to the case of the collision of a low-energy electron
with atomic hydrogen, \textit{i.e.} generating the wavefunctions needed for DWBA
calculations of electron impact ionization of atomic hydrogen. In this case
the static potential is given by:
\begin{displaymath}
V_{1 s} ( r ) = - 2 \left( 1 + \frac{1}{r} \right) \exp ( - 2 r )
\end{displaymath}
Like McDowell and collaborators, we use a Callaway-Temkin polarization
potential (see Drachman and Temkin \cite{Drachman}) which takes the form
\begin{displaymath}
V_{pol} ( r ) = - {\frac{9}{2 r^4}} \left[ 1 - e^{- 2 r} \left( 1 + 2
r + 2 r^2 + \frac{4}{3} r^3 + \frac{2}{3} r^4 + \frac{4}{27} r^5 \right) 
\right]
\end{displaymath}
To integrate the coupled equation systems (7) and (9) preference is given in
the present work to the "integral superposition" (I.S.) method which was
shown by Kobeissi \textit{et al} \cite{Kobeissi88, Kobeissi91b, Fakhreddine99}   to be highly accurate in the case of both single and
coupled differential equations. This requires potentials to be expressed in
analytical form. However, numerical potentials can generally be fitted by
analytical functions, for instance using cubic splines. The integration can
be safely taken out to very large radius, where ${\cal V}_{1 s} ( r)$
assumes its asymptotic form and the phase shift is determined by
equation (17). We refer to this method as Kobeissi-Fakhreddine-Tweed Exact
Exchange (KFTEE). 

Numerov integration of the coupled equation system (2) starting from series
solutions at the origin will be referred to as McDowell-Morgan-Myerscough
(McDMM) athough the code used differs from that of McDowell \textit{et al} \cite{McDowella} 
by the use of the Klapish-Robaux JWKB code as soon as the non-local exchange
term is negligible. For the sake of comparison, we have also made
calculations in which exact exchange is replaced by the use of the local
equivalent-exchange potentials of Furness and McCarthy \cite{Furness} or of Bransden
and Noble \cite{Bransden}. These are referred to respectively as Furness-McCarthy
Local Exchange (FMcCLE) or Bransden-Noble Local Exchange (BNLE). These
require the solution of a single differential equation, rather than an
integro-differential equation or a pair of coupled equations. We use Numerov
integration starting from a series solution at the origin and continuation by
the Klapish-Robaux JWKB solution from any convenient radius after the first
point of inflexion. The code used is similar to our version of the DWPO code
and we deliberately choose to switch to the JWKB solution at the same radius
as in the McDMM calculations. We use a regular radial mesh with steps of
$h$ for $0 \leq r \leq 1.2$, $2 h$ for $1.2 < r
\leq 4.8$, $4 h$ for $4.8 < r \leq 40.8$
a.u., the changeover to the JWKB solution being made at a
radius $r > 4.8$ determined by checking the matching in the DWPO case.
(For the purposes of collision calculations, where we determine the tails of
certain integrals by a method based on the second order JWKB solution, we
continue the radial mesh with a step of $8 h$ out to $r = 184.8$ a.u.) 
Calculations were made for three values of $h$: 0.004, 0.006 and 0.008
a.u. Whereas the phaseshifts are steplength-independent in the case of the
Local Exchange calculations, this is not so for the McDMM calculations at low
momentum. Results are given in tables 1 to 4 where they are compared to those
obtained with the present new KFTEE code. 

\begin{table}
\begin{tabular}{|c|c|c|c|c|c|c|c|c|} \hline & & S = 0 & & & & S = 1 & & \\
\hline $k$ & $h = .004$ & $h = .006$ & $h = .008$ & KFTEE &
 $h = .004$ & $h = .006$ & $h = .008$ & KFTEE \\ \hline 0.1 &
1.138750 & 1.134672 & 1.171174 & 2.527441 & 2.944466 & 2.944487 & 2.944556 &
2.948757\\ \hline 0.2 & 1.996521 & 1.995936 & 1.996479 & 2.034071 & 2.735678
& 2.735645 & 2.735678 & 2.735060\\ \hline 0.3 & 1.649999 & 1.650124 &
1.650246 & 1.665189 & 2.527570 & 2.527588 & 2.527605 & 2.523228\\ \hline 0.4
& 1.372797 & 1.372837 & 1.372796 & 1.384975 & 2.329332 & 2.329341 & 2.329332
& 2.322439\\ \hline 0.5 & 1.157391 & 1.157409 & 1.157391 & 1.168257 &
2.146210 & 2.146215 & 2.146210 & 2.137332\\ \hline 0.6 & 0.991071 & 0.991079
& 0.991087 & 1.000723 & 1.980222 & 1.980225 & 1.980228 & 1.969819\\ \hline
0.7 & 0.865011 & 0.865016 & 0.865020 & 0.873758 & 1.831592 & 1.831594 &
1.831596 & 1.819917\\ \hline 0.8 & 0.772639 & 0.772644 & 0.772644 & 0.779612
& 1.699554 & 1.699556 & 1.699556 & 1.743484\\ \hline 0.9 & 0.708203 &
0.708203 & 0.708199 & 0.713415 & 1.582765 & 1.582765 & 1.582763 & 1.621901\\
\hline 1.0 & 0.666187 & 0.666189 & 0.666178 & 0.670122 & 1.479626 & 1.479627
& 1.479620 & 1.507213\\ \hline 1.1 & 0.641285 & 0.641284 & 0.641271 &
0.644246 & 1.388498 & 1.388498 & 1.388489 & 1.407830\\ \hline 1.2 & 0.628568
& 0.628567 & 0.628552 & 0.630856 & 1.307821 & 1.307820 & 1.307809 &
1.320019\\ \hline 1.3 & 0.623787 & 0.623786 & 0.623773 & 0.625395 & 1.236182
& 1.236181 & 1.236170 & 1.242529\\ \hline 1.4 & 0.623565 & 0.623563 &
0.623551 & 0.624628 & 1.172346 & 1.172343 & 1.172333 & 1.174116\\ \hline 1.5
& 0.625441 & 0.625439 & 0.625429 & 0.626161 & 1.115244 & 1.115242 & 1.115232
& 1.113588\\ \hline
\end{tabular}
\caption{ Phase-shifts (rad.) for $l' = 0$ calculated as a function of $k$ (a.u) using the
McDMM code with three different steplengths $h$ (a.u.), compared to those obtained using
the KFTEE code. }
\end{table}

For $l' = 0$ (table 1) the McDMM code exhibits severe steplength
dependence for $k$ up to about 0.5 a.u. (3.4 eV energy) for both singlet
and triplet spin states. We also found that the singlet calculations for
$k$ up to about 0.3 a.u. were sensitive to the extent of the mesh
ranges chosen for the different steplength multiples. From $k = 0.6$
a.u. on the results appear to be stable to four decimal places. However,
the phase-shifts differ from those obtained from the KFTEE code: at $k =
0.6$ a.u. the difference is about $0.01$ rad. ; by $k
= 1.5$ a.u. it has fallen to $- 0.0008$ rad. for singlet and
$+ 0.0017$ rad. for triplet states, which in both cases comes to about
$0.1 \%$  error. This is probably due to the different ways in which the
phaseshifts are determined. In the KFTEE code we know both $f_1 ( r )$
and $f_1 ' ( r )$ and may use equation (14) to get $\tan \delta_
{l'}$ from their values at a single mesh-point. In the McDMM code
we only dispose of $f_1 ( r )$ and so have to use its values at two
mesh-points to get $\tan \delta_{l'}$ ; this procedure is
more subject to numerical error, even if care is taken to choose points
separated by a half-period of the JWKB phase function. 

\begin{table}
\begin{tabular}{|c|c|c|c|c|c|c|c|c|} \hline & & S = 0 & & & & S = 1 & & \\
\hline $k$ & $h = .004$ & $h = .006$ & $h = .008$ & KFTEE &
 $h = .004$ & $h = .006$ & $h = .008$ & KFTEE \\ \hline 0.1 &
0.006806 & 0.006806 & 0.006805 & 0.006873 & 0.011107 & 0.011107 & 0.011105 &
0.011161\\ \hline 0.2 & 0.018017 & 0.018017 & 0.018016 & 0.018043 & 0.050578
& 0.050577 & 0.050576 & 0.050605\\ \hline 0.3 & 0.023633 & 0.023633 &
0.023633 & 0.023657 & 0.121599 & 0.121599 & 0.121599 & 0.121611\\ \hline 0.4
& 0.021210 & 0.021210 & 0.021209 & 0.021230 & 0.215073 & 0.215072 & 0.215072
& 0.215060\\ \hline 0.5 & 0.013588 & 0.013588 & 0.013588 & 0.013606 &
0.311177 & 0.311177 & 0.311176 & 0.311150\\ \hline 0.6 & 0.005301 & 0.005301
& 0.005301 & 0.005314 & 0.391342 & 0.391342 & 0.391342 & 0.391291\\ \hline
0.7 & 0.000175 & 0.000175 & 0.000175 & 0.000185 & 0.447613 & 0.447613 &
0.447613 & 0.447559\\ \hline 0.8 & 0.000478 & 0.000478 & 0.000478 & 0.000486
& 0.481454 & 0.481454 & 0.481453 & 0.481395\\ \hline 0.9 & 0.006922 &
0.006922 & 0.006922 & 0.006933 & 0.498186 & 0.498186 & 0.498186 & 0.498122\\
\hline 1.0 & 0.019049 & 0.019049 & 0.019049 & 0.019062 & 0.503268 & 0.503268
& 0.503268 & 0.503206\\ \hline
\end{tabular}
\caption{ Phase-shifts (rad.) for $l' = 1$ calculated using the
McDMM code with three different steplengths, compared to those obtained using
the KFTEE code. }
\end{table}

For $l' = 1$ (table 2) the steplength dependence of the McDMM code is
less important. This is also the case for $l' \geq 2$ so in tables
3 and 4 we give McDMM results for $h = 0.006$ a.u. only. We again
find that the McDMM phaseshifts differ from those calculated by the present
KFTEE method. But the differences are much smaller than in the case of $l' =
0$: generally, from $k = 0.6$ a.u. onwards, only the fourth decimal
changes. Presumably, the inhomogeneous solution is more sensitive than
the homogeneous one to the use of Numerov integration. Instabilities may also
arise in the determination of $A ( k )$, which depends on calculating
short-range integrals for the overlap of the target wavefunction with
solutions of the homogeneous and of the inhomogeneous equations. These integrals
are sensitive to the behaviour of the solutions at small radius, so we would
not expect them to be affected by errors which accumulate as the solution is
integrated outwards. Even if the solutions become unstable at long range, $A
( k )$ should not be badly affected. So the errors and instabilities in
the McDMM code for small $k$ (energies below $\sim 5$ eV) can be
imputed to the use of Numerov integration in a coupled equation problem and
will presumably get worse as the number of equations increases. Hence the
usefulness of the present Canonical Functions method in the case of a target
with outer electron orbitals of angular momentum $l > 0$, which requires 
the solution of a system of many more 
coupled equations than only the two needed for hydrogenic atoms. 

\begin{table}
\begin{tabular}{|c|c|c|c|c|c|c|c|c|} \hline & $l' = 2$ & $l' = 2$ & $l' = 3$ & $l' = 3$
& $l' = 4$ & $l' = 4$ & $l' = 5$ & $l' = 5$ \\ \hline $k$ & McDMM
& KFTEE & McDMM & KFTEE & McDMM &
 KFTEE & McDMM & KFTEE\\ \hline 0.1 &
0.001287 & 0.001344 & 0.000334 & 0.000449 & unstable & 0.000204 &
unstable & 0.000110\\ \hline 0.2 & 0.005231 & 0.005269 & 0.001765
& 0.001795 & 0.000776 & 0.000816 & 0.000401 & 0.000439\\ \hline 0.3 &
0.011215 & 0.011234 & 0.004005 & 0.004028 & 0.001817 & 0.001837 & 0.000962 &
0.000988\\ \hline 0.4 & 0.018215 & 0.018227 & 0.007071 & 0.007085 & 0.003247
& 0.003264 & 0.001743 & 0.001758\\ \hline 0.5 & 0.025156 & 0.025163 &
0.010797 & 0.010806 & 0.005074 & 0.005084 & 0.002733 & 0.002745\\ \hline 0.6
& 0.031323 & 0.031322 & 0.014959 & 0.014962 & 0.007255 & 0.007263 & 0.003942
& 0.003949\\ \hline 0.7 & 0.036537 & 0.036534 & 0.019299 & 0.019301 &
0.009737 & 0.009741 & 0.005354 & 0.005360\\ \hline 0.8 & 0.041023 & 0.041016
& 0.023616 & 0.023610 & 0.012437 & 0.012436 & 0.006953 & 0.006958\\ \hline
0.9 & 0.045182 & 0.045174 & 0.027784 & 0.027781 & 0.015271 & 0.015269 &
0.008710 & 0.008711\\ \hline 1.0 & 0.049394 & 0.049382 & 0.031763 & 0.031756
& 0.018163 & 0.018158 & 0.010593 & 0.010591\\ \hline 
\end{tabular}
\caption{ Phase-shifts (rad.) for $S = 0$ and $l' \geq
2$, calculated using the McDMM code with a steplength of $h = 0.006$,
compared to those obtained using the KFTEE code. }
\end{table}

\begin{table}
\begin{tabular}{|c|c|c|c|c|c|c|c|c|} \hline & $l' = 2$ & $l' = 2$ & $l' = 3$ & $l' = 3$
& $l' = 4$ & $l' = 4$ & $l' = 5$ & $l' = 5$ \\ \hline $k$ & McDMM
& KFTEE & McDMM & KFTEE & McDMM &
 KFTEE & McDMM & KFTEE\\ \hline 0.1 &
0.001295 & 0.001358 & 0.000334 & 0.000449 & unstable & 0.000204 &
unstable & 0.000110\\ \hline 0.2 & 0.005456 & 0.005492 & 0.001768
& 0.001798 & 0.000776 & 0.000816 & 0.000401 & 0.000439\\ \hline 0.3 &
0.012687 & 0.012706 & 0.004035 & 0.004059 & 0.001818 & 0.001837 & 0.000962 &
0.000988\\ \hline 0.4 & 0.023298 & 0.023310 & 0.007249 & 0.007262 & 0.003254
& 0.003270 & 0.001743 & 0.001758\\ \hline 0.5 & 0.037315 & 0.037320 &
0.011437 & 0.011446 & 0.005110 & 0.005120 & 0.002735 & 0.002748\\ \hline 0.6
& 0.054197 & 0.054200 & 0.016625 & 0.016628 & 0.007384 & 0.007392 & 0.003953
& 0.003961\\ \hline 0.7 & 0.072899 & 0.072899 & 0.022758 & 0.022760 &
0.010084 & 0.010088 & 0.005389 & 0.005396\\ \hline 0.8 & 0.092140 & 0.092132
& 0.029699 & 0.029696 & 0.013194 & 0.013197 & 0.007049 & 0.007053\\ \hline
0.9 & 0.110743 & 0.110721 & 0.037226 & 0.037223 & 0.016684 & 0.016684 &
0.008925 & 0.008928\\ \hline 1.0 & 0.127837 & 0.127824 & 0.044079 & 0.045069
& 0.020494 & 0.020494 & 0.011006 & 0.011007\\ \hline 
\end{tabular}
\caption{ Phase-shifts (rad.) for $S = 1$ and $l' \geq
2$, calculated using the McDMM code with a steplength of $h = 0.006$,
compared to those obtained using the KFTEE code. }
\end{table}

In tables 3 (singlet case) and 4 (triplet case) we compare phaseshifts from
the McDMM and KFTEE codes for partial waves $l' = 2 , 3 , 4 , 5$. We see
that at $k = 0.1$ the McDMM code is either unstable or else fails to get
even the first significant figure correct. Agreement to two signifiacant
figures at all $l'$ is only obtained for $k > 0.5$. Agreement to
three significant figures is obtained from $k \approx 0.7$ a.u. (7 eV
energy) onwards. At this energy it is necessary to include about ten
partial waves in collision calculations and slight changes in the value of
the phaseshift can significantly affect the cross section values even though
the radial integrals (which depend mainly on the short range behaviour of the
wavefunctions) show hardly any differences.

Finally, in figures 1 to 4, we compare, for the lowest two partial waves, the
phaseshifts obtained using the KFTEE code with those from the local
equivalent-exchange potential models FMcCLE and BNLE. Important differences
are found up to $k \approx 1.1$ a.u. (15 eV
energy). In the case of singlet spin states, for $l' = 0$
(figure 1) the equivalent-exchange models give severe underestimates but for
$l' = 1$ (figure 3) they are acceptble. In the case of triplet spin
states, for $l' = 0$ (figure 2) results are quite good but for $l' =
1$ the BNLE model gives a severe overestimate. For higher partial
waves equivalent-exchange models give fairly satisfactory results. Since
Distorted Wave Born collision calculations near to threshold converge with
only about five partial waves, we would hesitate then to use a local
equivalent-exchange potential. 

\begin{figure}[htbp]
\begin{center}
\scalebox{1.0}{\includegraphics[angle=-90]{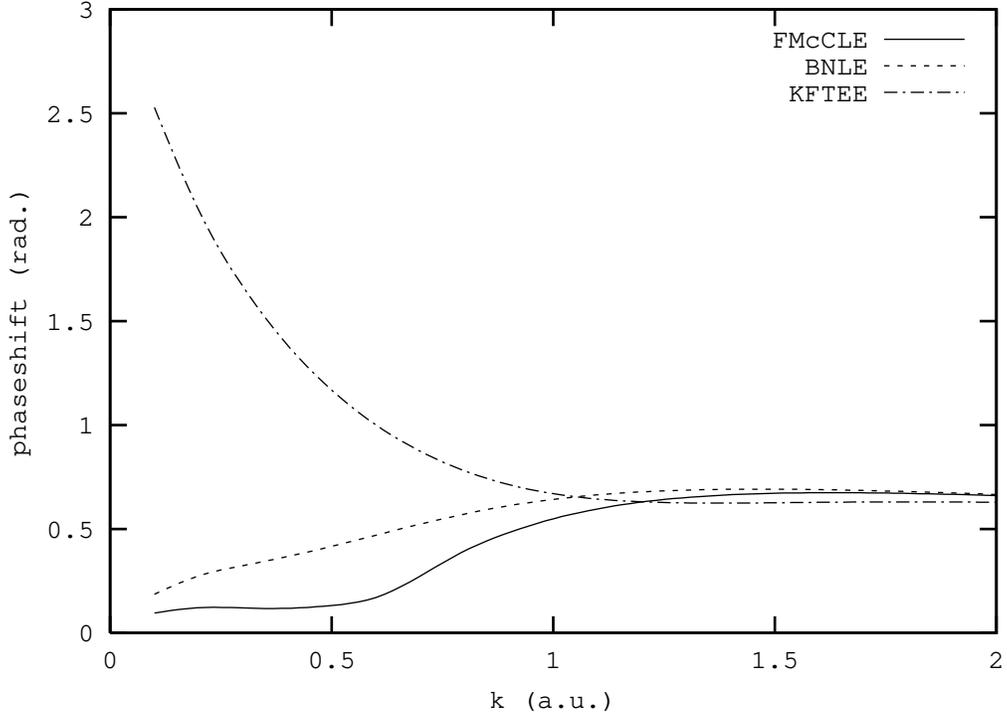}}
\caption{Singlet state phase-shifts for $l' = 0$ as a function
of impulsion $k$. Full line: present KFTEE non-local exchange model;
dashed line: BNLE equivalent-exchange potential model; dot-dash line: FMcCLE
equivalent-exchange potential model.}
\end{center}
\end{figure}

\begin{figure}[htbp]
\begin{center}
\scalebox{1.0}{\includegraphics[angle=-90]{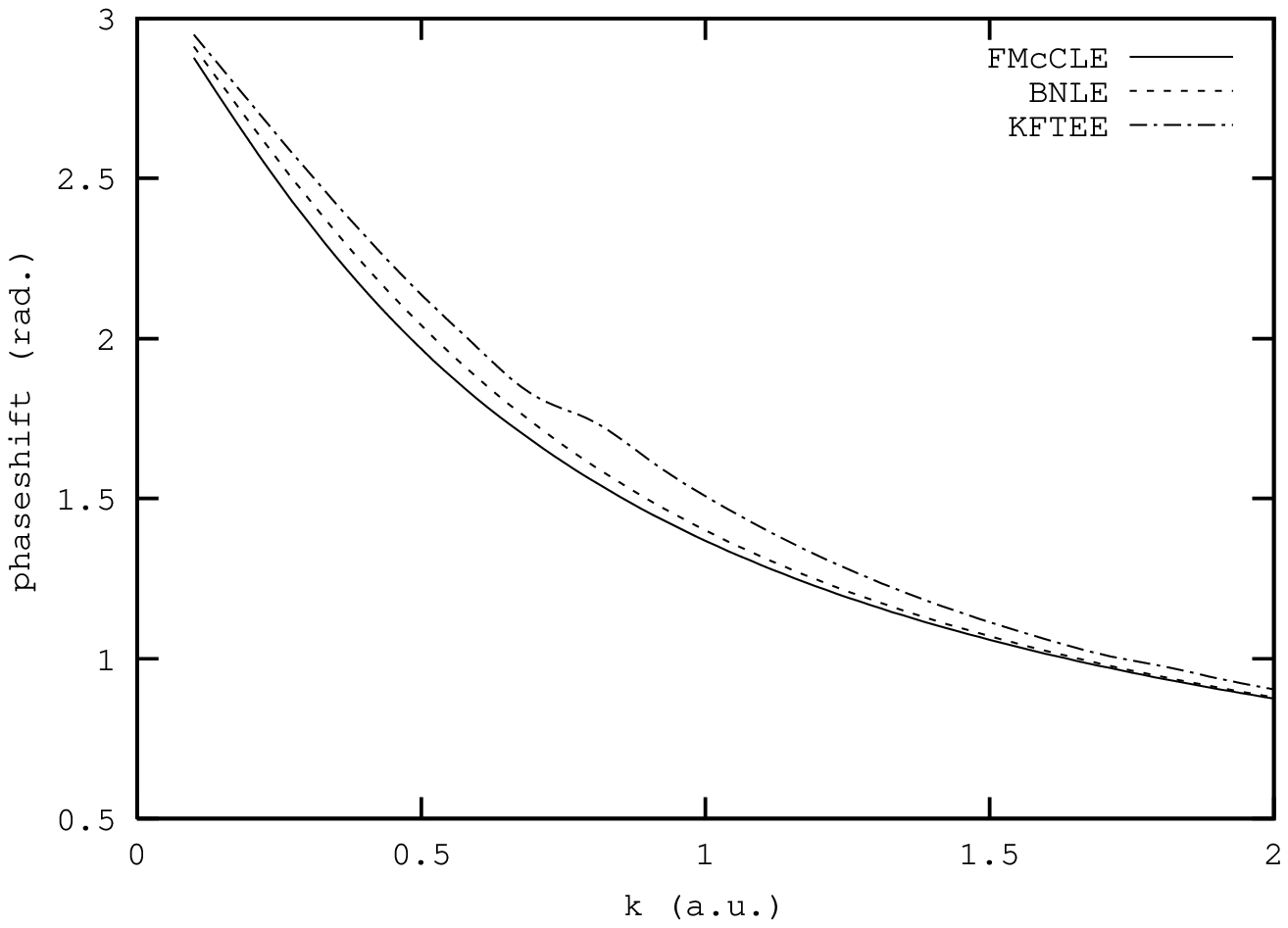}}
\caption{Triplet state phase-shifts for $l' = 0$. As figure 1.}
\end{center}
\end{figure}

\begin{figure}[htbp]
\begin{center}
\scalebox{1.0}{\includegraphics[angle=-90]{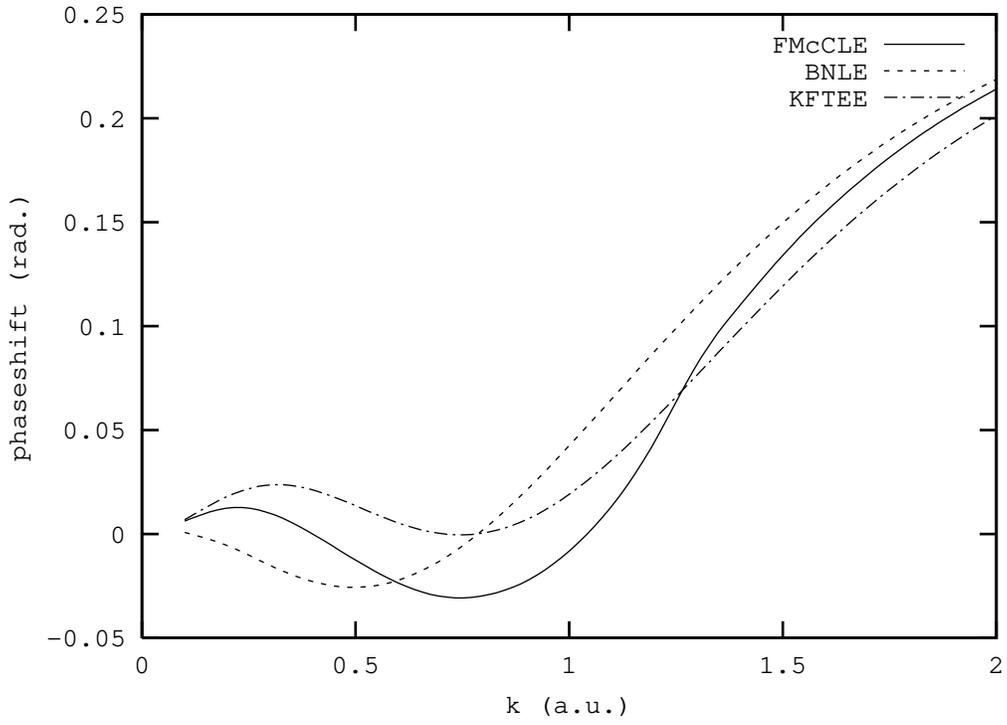}}
\caption{Singlet state phase-shifts for $l' = 1$. As figure 1.}
\end{center} 
\end{figure}

\begin{figure}[htbp]
\begin{center}
\scalebox{1.0}{\includegraphics[angle=-90]{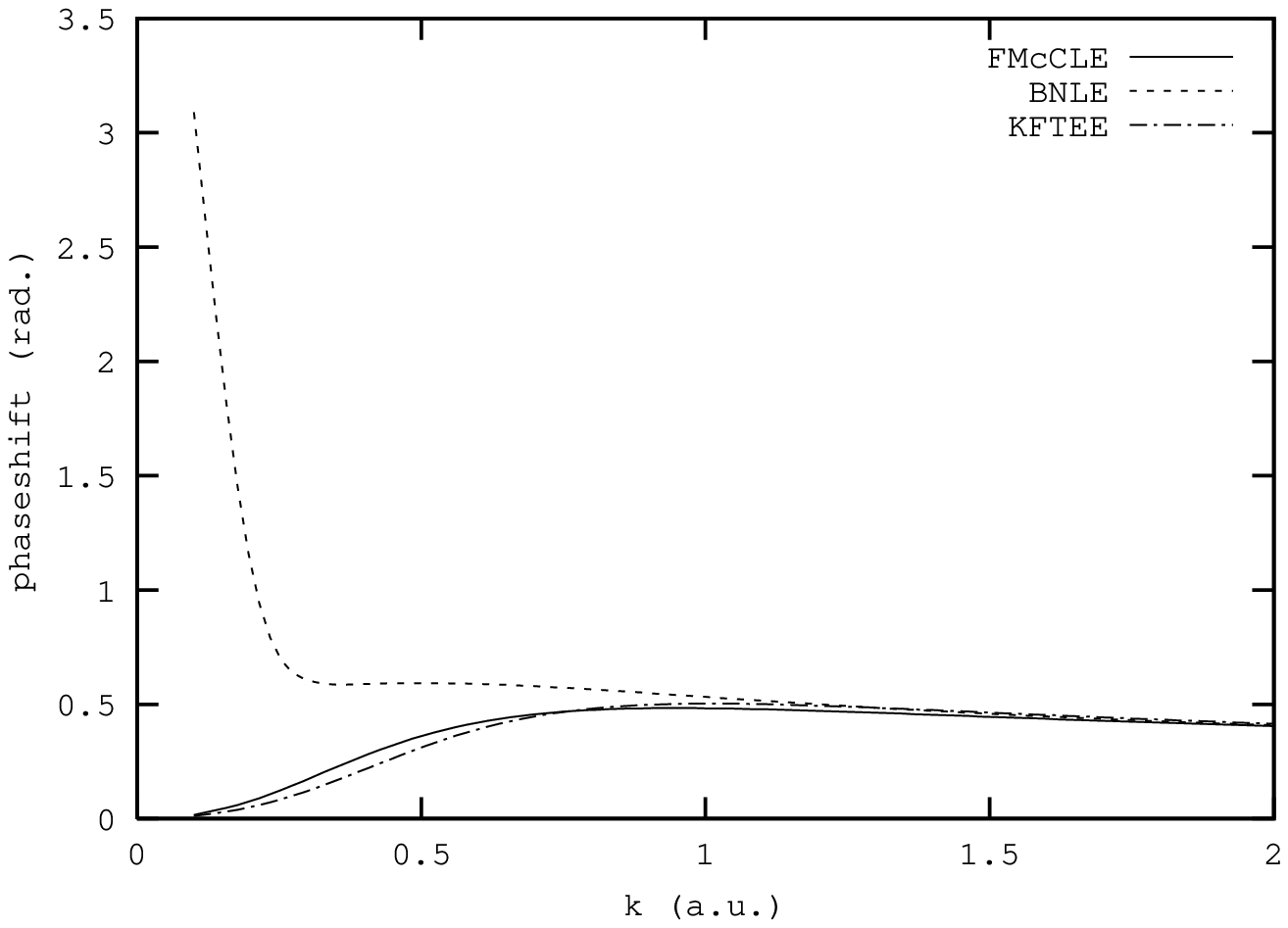}}
\caption{Triplet state phase-shifts for $l' = 1$. As figure 1.}
\end{center}
\end{figure}

\section{Conclusions and Prospects}

We have shown that a distorted wave code, with exact treatment of exchange but
using Numerov integration, breaks down through numerical instabilities at
energies below about 5 eV. We have also shown that codes with local
equivalent-exchange potentiels give poor results for low partial waves at
energies below 15 eV. We propose an alternative method treating exchange
exactly but using the Canonical Function technique to integrate the coupled
equations. This we believe to be numerically stable even at extremely small
energies. The present work is the first step towards a general code
applicable to target atoms or molecules in any angular momentum state and
capable of using numerically generated potentials. An alternative code is
under development in which solutions are obtained by Runge-Kutta integration
on a regular grid out to a radius where exchange terms are negligible; 
phase-shifts are obtained by
comparison with the Klapisch-Robaux iterative JWKB code \cite{bar}.
In this code we have on the one hand $f_1 ( r )$ and $f_1 ' ( r )$
and on the other a phase function, its derivative and a slowly-varying
amplitude function for which we can safely generate the derivative
numerically.  So we can get $\tan \delta_{l'}$ using a single
mesh point, but without the need to carry numerical integration of the
coupled equations out to a radius (very big for small $k$) where the
asymptotic behaviour is given by $s_{l'} ( k r )$ and $c_{l'} ( k r
)$ . The new code will give us the opportunity to check the relative
accuracy of phaseshift determinations using respectively $f_1 ( r )$
and $ f_1 ' ( r )$ at one mesh-point or $f_1 ( r )$ at
two. We also intend to test the effects at low energy of using Bethe-Reeh
type multipole polarisation potentials \cite{Drachman} which we
will generate numerically, for instance in the case of \textit{Na} and the rare 
gases; this will enable us to extend the work of Rouet \textit{et al} \cite{Rouet}. We are
particularly interested in applications to $( e , 2 e )$ and $( \gamma ,
2 e )$ processes involving polarised electrons.

\section*{Acknowledgments}

This work was partly financed by the Lebanese University and the C.N.R.S.L.
K.F. thanks the Universit\'e de Bretagne Occidentale for a two month stay as
Professeur Invit\'e and the L.C.E.A. for its hospitality.

\end{document}